\documentclass[a4paper,10pt]{article}

\usepackage[utf8]{inputenc}
\usepackage[T1]{fontenc}
\usepackage[frenchb,english]{babel}
\usepackage{textcomp}
\usepackage{amsmath}
\usepackage{amsfonts}
\usepackage{amssymb}
\usepackage{graphicx}
\usepackage{vmargin}
\usepackage[bookmarks=true,bookmarksnumbered=true,colorlinks=false,pdfborder={0 0 0}]{hyperref}
\usepackage{fancyhdr}
\usepackage{lastpage}
\usepackage{natbib}
\usepackage{cases}
\usepackage[normal]{subfigure}
\usepackage{upgreek}

\setmarginsrb{25mm}{20mm}{25mm}{15mm}{12pt}{5mm}{0pt}{11mm}

\begin{document}


\pagestyle{fancy}
\chead{}
\lhead{\textit{B. Lenoir et al., 2013}}
\rhead{\textit{\thepage/\pageref{LastPage}}}
\cfoot{}
\renewcommand{\headrulewidth}{0pt}
\renewcommand{\footrulewidth}{0pt}



\title{Unbiased acceleration measurements with an electrostatic accelerometer on a rotating platform}
\author{Benjamin Lenoir\textsuperscript{a}, Bruno Christophe\textsuperscript{a}, Serge Reynaud\textsuperscript{b} \\ \small \textsuperscript{a} \textit{Onera -- The French Aerospace Lab, 29 avenue de la Division Leclerc, F-92322 Ch\^atillon, France} \\ \small \textsuperscript{b} \textit{Laboratoire Kastler Brossel (LKB), ENS, UPMC, CNRS, Campus Jussieu, F-75252 Paris Cedex 05, France} \\ \\ \small Published in \textit{Advances in Space Research} 51 (2012) 188-197 \\ \small doi: \href{http://dx.doi.org/10.1016/j.asr.2012.08.012}{10.1016/j.asr.2012.08.012}}
\date{6 January 2013}
\maketitle



\begin{abstract}
The Gravity Advanced Package is an instrument composed of an electrostatic accelerometer called MicroSTAR and a rotating platform called Bias Rejection System. It aims at measuring with no bias the non-gravitational acceleration of a spacecraft. It is envisioned to be embarked on an interplanetary spacecraft as a tool to test the laws of gravitation.

MicroSTAR is based on Onera's experience and inherits in~orbit technology. The addition of the rotating platform is a technological upgrade which allows using an electrostatic accelerometer to make measurements at low frequencies with no bias. To do so, the Bias Rejection System rotates MicroSTAR such that the signal of interest is separated from the bias of the instrument in the frequency domain. Making these unbiased low-frequency measurements requires post-processing the data. The signal processing technique developed for this purpose is the focus of this article. It allows giving the conditions under which the bias is completely removed from the signal of interest. And the precision of the unbiased measurements can be fully characterized: given the characteristics of the subsystems, it is possible to reach a precision of 1~pm~s$^{-2}$ on the non-gravitational acceleration for an integration time of 3~h.


\end{abstract}



\section{Introduction}\label{section:introduction}

The experimental tests of gravitation are in good agreement with its current theoretical formulation referred to as General Relativity \citep{will2006confrontation}. But contrary to the quantum description of the three other fundamental interactions, it is a classical theory, which suggests that another description of gravitation lies beyond General Relativity. From the experimental point of view, there are still open windows for deviations from General Relativity at short range \citep{adelberger2003tests} and at long range \citep{jaekel2005gravity}.
Galactic and cosmic observations also challenge General Relativity. The rotation curves of galaxies and the relation between redshifts and luminosities of supernovae, which are interpreted as manifestations of ``dark matter'' and ``dark energy'' respectively \citep{frieman2008dark,bertone2005particle}, may also be seen as a hint that General Relativity could be an imperfect description of description at these large scales \citep{aguirre2001astrophysical,nojiri2007introduction}.

In this context, testing General Relativity at the largest possible scales is essential. For man-made instruments, the Solar System can be used as a laboratory for gravitational experiments. NASA performed such a test of gravitation with the Pioneer 10 and 11 missions. The outcome was a signal now known as the Pioneer anomaly \citep{anderson1998indication,anderson2002study,turyshev2011support}. The long term variations of this signal may be explained as an anisotropic
thermal effect \citep{bertolami2010estimating,rievers2010modeling,rievers2011high,turyshev2012support} but periodic anomalies have also been identified \citep{levy2009pioneer,courty2010simulation}. Nevertheless, the \textit{Roadmap for Fundamental Physics in Space} issued by ESA in 2010 \citep{esa2010roadmap} stresses the importance of testing gravitation at large scales with missions to the outer planets.
To do so, it recommends the development of accelerometers compatible with spacecraft tracking at the 10 pm s$^{-2}$ level.

Several missions have already been proposed \citep{anderson2002mission,dittus2005mission,johann2008exploring,bertolami2007mission,christophe2009odyssey,wolf2009quantum} with the aim of improving the knowledge of the gravitational field in the Solar System. Many of them propose to embark an accelerometer which will measure the non-gravitational forces acting on the spacecraft in order to distinguish unambiguously the non-gravitational accelerations from gravitational effects. In practice, accelerometers measure a combination of the spacecraft non-gravitational acceleration and additional terms \citep{carbone2005characterization,carbone2007thermal}.
It will be assumed in this article that these additional terms either are negligible or can be corrected, such that the external signal will be referred to as the spacecraft non-gravitational acceleration \citep{lenoir2011electrostatic}.

This article deals with the Gravity Advanced Package, an instrument proposed on Laplace mission \citep{biesbroek2008laplace}, which is designed to make measurements of the non-gravitational acceleration of a spacecraft with no bias. It provides an additional observable which measures the departure of the spacecraft from geodesic motion. To do so, post-processing is required, which is the subject of this article.
It allows retrieving separately the acceleration without bias and the bias of the instrument, these quantities being referred to as ``post-processed'' quantities.

In a first part, the Gravity Advanced Package will be presented as well as its performances and the measurement principle. Then the post-processing method will be described and conditions will be given so that the bias can be effectively removed from the measurement. These conditions will allow deriving measurement procedures to make unbiased measurements. The emphasis will then be put on the characterization of the post-processed quantities, i.e. the quantities without bias, and on their uncertainty.

This paper is focused on the performances of the accelerometer. The constraints in the integration of the instrument in the spacecraft with the aim of preserving these performances requires additional work. The OSS mission \citep{christophe2011ossExpAstro} proposes a spacecraft design which takes into account these concerns.

\section{Instrument principle, design and performance}\label{section:instrument}

The Gravity Advanced Package is made of an electrostatic accelerometer, called MicroSTAR, which can be rotated with the Bias Rejection System. This technological upgrade allows removing the bias introduced by MicroSTAR.

\subsection{Overview of the instrument}

MicroSTAR is a 3-axes electrostatic accelerometer \citep{josselin1999capacitive} based on ONERA's expertise in this field \citep{touboul1999accelerometers,hudson2007development,touboul2001microscope}. In orbit technology (CHAMP, GRACE and GOCE missions) is used with improvements to reduce power consumption, size and mass.

The core of the accelerometer is composed of a proof mass inside a cage made of six identical plates. The motion of the proof mass with respect to the cage is detected by capacitive measurement. A control loop adjusts the potentials of the electrodes in order to keep the proof mass at the center of the cage. The numerical values of these potentials, which are the outputs of the instrument, are proportional to the components of the acceleration of the proof mass with respect to the cage on each axis of the accelerometer.

The Bias Rejection System is a rotating platform composed of a rotating actuator and a high resolution angular encoder working in closed loop operation. Piezo-electric technology is envisioned for the actuator. It allows designing a device with no need for any kind of gear, so as to reduce mass and volume. The piezo-electric motor is operated in a slip-stick mode.
Finally, even if piezo-electric motors have a non-zero torque in the power-off mode, a blocking system will be implemented to prevent unwanted motion during launch and maneuvers.

\subsection{Performance of the electrostatic accelerometer}

The performance of MicroSTAR is measured via the power spectrum density of the noise on the measured acceleration \cite[Fig. 4]{lenoir2011electrostatic}. The analytic formula $S_{n}$ (in m$^2$~s$^{-4}$~Hz$^{-1}$) as a function of frequency, for a measurement range equal to $1.8 \times 10^{-4}$~m~s$^{-2}$, is
\begin{equation}
 \sqrt{S_{n}(f)} = K \sqrt{ 1 + \left(\frac{f}{4.2 \ \mathrm{mHz}}\right)^{-1} + \left(\frac{f}{0.27 \ \mathrm{Hz}}\right)^{4} }
 \label{eq:acc_noise}
\end{equation}
with $K = 5.7 \times 10^{-11} \ \mathrm{m}.\mathrm{s}^{-2}.\mathrm{Hz}^{-1/2}$.

In addition, the instrument has a bias. It corresponds to the deterministic low-frequency variations of the accuracy of the accelerometer. It is due to the gold wires which are used to keep the polarization of the proof mass constant and to the geometrical imperfections of the instrument. In previous missions relying on electrostatic accelerometers (CHAMP, GRACE and GOCE), this bias was not a problem since the measurement bandwidth was 0.1--100~mHz. On the contrary, for the application foreseen in this article, very low-frequencies measurements are to be made and it is therefore required to remove the bias from the measurements.

\subsection{Measurement principle}

The measurements made by MicroSTAR along its three axes $x$, $y$ and $z$, which are supposed to be orthogonal\footnote{The orthogonality of the measurement axes depends on the orthogonality of the proof mass faces. Assuming a non-gravitational acceleration equal to $10^{-7}$~m~s$^{-2}$ (cf. section \ref{subsection:generalized_noise}), the orthogonality needs to be controlled at a level of 10~$\upmu$rad in order to have an accuracy on the measurement of 1~pm~s$^{-2}$. This level is technologically reachable.}, are
\begin{equation}
 \begin{bmatrix}
  m_x \\
  m_y \\
  m_z
 \end{bmatrix}
 =
 \begin{bmatrix}
  (1+\delta k_{1x})a_x + k_{2x} {a_x}^2 + b_x + n_x \\
  (1+\delta k_{1y})a_y + k_{2y} {a_y}^2 + b_y + n_y \\
  (1+\delta k_{1z})a_z + k_{2z} {a_z}^2 + b_z + n_z
 \end{bmatrix}
 \label{eq:mes_1}
\end{equation}
where $\delta k_{1\kappa}$, $k_{2\kappa}$, $b_\kappa$ and $n_\kappa$ ($\kappa\in \{x;y;z\}$) are respectively the scale factors, the quadratic factors, the bias and the noise on each axis. The bias has a deterministic temporal variation whereas the noise is a null-mean stationary stochastic process whose PSD is given by Eq.~\eqref{eq:acc_noise}. The quantities $a_\kappa$ are the components of the non-gravitational acceleration in the reference frame of MicroSTAR. As far as orbit reconstruction is concerned, these quantities are not the ones of interest since the Bias Rejection System rotates MicroSTAR with respect to the spacecraft. The spacecraft is supposed to be stabilized along the three axes. The transformation matrix $P$ moves a vector from the spacecraft reference frame (whose axes are $X$, $Y$ and $Z$) to the accelerometer reference frame. In its simplest form (but without any loss of generality), the expression of $P$ is
\begin{equation}
 P =
 \begin{bmatrix}
  1 & 0 & 0 \\
  0 & \cos(\theta) & \sin(\theta) \\
  0 & -\sin(\theta) & \cos(\theta)
 \end{bmatrix}
\end{equation}
where $\theta$ is a monitored angle, which measures the rotation of the accelerometer with respect to the spacecraft. Considering only the plane perpendicular to the axis of rotation of the accelerometer, Eq.~\eqref{eq:mes_1} becomes
\begin{subequations}
 \begin{numcases}{}
  m_y = (1+\delta k_{1y}) \left[\cos(\theta) a_Y  + \sin(\theta) a_Z \right] \nonumber \\ + k_{2y} \left[\cos(\theta) a_Y + \sin(\theta) a_Z \right]^2 + b_y + n_y \\
  m_z = (1+\delta k_{1z}) \left[-\sin(\theta) a_Y + \cos(\theta) a_Z \right] \nonumber \\ + k_{2z} \left[-\sin(\theta) a_Y + \cos(\theta) a_Z \right]^2 + b_z + n_z
 \end{numcases}
 \label{eq:pos_pb}
\end{subequations}
The measurements on the axes $y$ and $z$ are combinations of the quantities $a_Y$ and $a_Z$. These are the quantities needed so as to measure the impact of non-gravitational forces on the trajectory of the spacecraft. This fact associated with the possibility to give the angle $\theta$ any possible time variation allows measuring $a_Y$ and $a_Z$ without bias. On the contrary, on the axis $x$, there is no possibility  with this instrument to remove the bias from the measurement $m_x$ so as to retrieve $a_X$. To do so, another rotating platform would be required. It is not the topic of this article but the method developed here can be applied to this more complex setup. In practice, the axes $Y$ and $Z$ will be in the orbit plane, in which non-gravitational forces are expected to impact the trajectory of the spacecraft.

In the following, it will be assumed that $N$ measurements are made with a sampling frequency called $f_s$. It corresponds to a time step called $\delta t = 1/f_s$. The scale and quadratic factors will be supposed to be constant. In Eq.~\eqref{eq:pos_pb}, there are therefore $4N$ unknowns ($a_Y$, $a_Z$, $b_y$, $b_z$ at each sampling time) and $2N$ measurements ($m_y$,$m_z$) spoiled by noise ($n_y$,$n_z$). In the rest of this article, for each of these eight quantities as well as for $\theta$, the notation $\mathbf{x}$ will be a vector of $\mathcal{M}_{N,1}(\mathbb{R})$ whose components are the values of $x$ at each sampling time and $x_k$ is the value of $x$ at the sampling time $k \times \delta t$.

\section{Signal processing method}\label{section:processing}

The relations between the measurements made by the instrument, $m_y$ and $m_z$, and the non-gravitational acceleration in the spacecraft reference frame, $a_Y$ and $a_Z$, has been expressed. In this section, the data processing method is presented. In particular, conditions are derived in order to remove the bias from the measurements.

\subsection{Linearization of the problem}

A first step is to linearize Eq.~\eqref{eq:pos_pb} such that they can be written in a matrix form. To do so, it is assumed that $k_{2y}=k_{2z}=0$. This hypothesis will be shown in paragraph \ref{subsection:choice_signal} not to be restrictive in the framework presented here. The two following diagonal matrices, belonging to $\mathcal{M}_N(\mathbb{R})$,
\begin{equation}
 \Lambda_c = \mathrm{diag}[\cos(\theta_k)]
 \ \ \textnormal{and} \ \
 \Lambda_s = \mathrm{diag}[\sin(\theta_k)], \ k\in||1;N||
\end{equation}
allow writing Eq.~\eqref{eq:pos_pb} in the matrix form
\begin{equation}
 M = J X + E
 \label{eq:linear_system}
\end{equation}
with
\begin{equation}
 X =
 \begin{bmatrix}
  \mathbf{a_Y} \\
  \mathbf{a_Z} \\
  \mathbf{b_y} \\
  \mathbf{b_z}
 \end{bmatrix}
 \ \textnormal{,} \ \
 M =
 \begin{bmatrix}
  \mathbf{m_y} \\
  \mathbf{m_z}
 \end{bmatrix}
 \ \textnormal{,} \ \
 E =
 \begin{bmatrix}
  \mathbf{n_y} \\
  \mathbf{n_z}
 \end{bmatrix}
\end{equation}
and
\begin{equation}
  J =
  \begin{bmatrix}
      (1+\delta k_{1y}) \Lambda_c & (1+\delta k_{1y})  \Lambda_s & \mathrm{Id}_N & 0 \\
    - (1+\delta k_{1z}) \Lambda_s & (1+\delta k_{1z}) \Lambda_c & 0 & \mathrm{Id}_N
  \end{bmatrix}.
\end{equation}
The set of solutions for this system is infinite. It is the affine space $X_p + \ker(J)$, where $X_p$ is a given solution of the linear equation. This formal resolution gives no useful information on the non-gravitational acceleration since it provides an infinite number of solutions.

\subsection{Generalized noise}\label{subsection:generalized_noise}

Before going further, it is necessary to consider the matrix $J$ more carefully. When solving Eq.~\eqref{eq:linear_system}, $J$ is supposed to be perfectly known. It is however not the case since the knowledge of the angle $\theta$ involved in the definition of $J$ may suffer a bias and noise. There is a discrepancy between the true value of the rotation angle $\theta^*$ and the measured one $\theta$:
\begin{equation}
  \theta = \theta^* + b_\theta + \delta\theta
\end{equation}
where $b_\theta$ is a bias and $\delta\theta$ a random process (whose mean value is equal to zero). Using the same notations, this leads to a noise described by $\delta\Lambda_c$ and $\delta\Lambda_s$ on the matrices $\Lambda_c$ and $\Lambda_s$\footnote{$\delta\Lambda_c = \mathrm{diag}[\cos(\theta_k + \delta\theta_k) - \cos(\theta_k)] \approx \mathrm{diag}[- \delta\theta_k \sin(\theta_k)]$ and $\delta\Lambda_s = \mathrm{diag}[\sin(\theta_k + \delta\theta_k) - \sin(\theta_k)] \approx \mathrm{diag}[\delta\theta_k \cos(\theta_k)]$}.

The impact of the bias on the precision of the measurement has been assessed in \citep{lenoir2011electrostatic} and it has been shown that $b_\theta$ must be smaller than $10^{-5}$ rad in order to meet the expected performances.

In order to take into account the impact of the noise $\delta\theta$, it is possible to introduce a generalized noise: the quantities $\mathbf{n_y}$ and $\mathbf{n_z}$ in equation \eqref{eq:linear_system} are replaced by $\mathbf{\tilde{n}_y} = \mathbf{n_y} + \mathbf{\hat{n}_y}$ and $\mathbf{\tilde{n}_z} = \mathbf{n_z} + \mathbf{\hat{n}_z}$ with
\begin{subequations}
  \begin{numcases}{}
    \mathbf{\hat{n}_y} = (1+\delta k_{1y}) \left[\delta\Lambda_c \mathbf{a_Y} + \delta\Lambda_s \mathbf{a_Z} \right] \\
    \mathbf{\hat{n}_z} = (1+\delta k_{1z}) \left[-\delta\Lambda_s \mathbf{a_Y} + \delta\Lambda_c \mathbf{a_Z} \right]
  \end{numcases}
  \label{eq:generalized_noise}
\end{subequations}
This additional noise depends on the non-gravitational accelerations $\mathbf{a_Y}$ and $\mathbf{a_Z}$ and on $\delta\Lambda_c$ and $\delta\Lambda_s$. As a result, the smaller the magnitude of the external acceleration is, the smaller the noise due to the uncertainty on $\theta$ is.

It can be used to derive the requirements on $\delta \theta$ such that the predominant source of uncertainty is MicroSTAR and not the Bias Rejections System. To have such a result, one needs $S_{\hat{n}}(f) \ll S_n(f)$ around the modulation frequency $1/\tau$ (cf. section \ref{subsection:choice_signal}), where $S_{\hat{n}}(f)$ is the power spectrum density of the noise $\hat{n}$ due to the rotating platform (cf. Eq.~\eqref{eq:generalized_noise}). Assuming that $a_y \approx a_z \approx a_{NG}$, we have $\hat{n} \approx \delta\theta a_{NG}$. This leads to the following requirement :
\begin{equation}
 \forall f\in \left[\frac{1}{2\tau} ; \frac{3}{2\tau} \right], S_{\delta \theta}(f) \ll \frac{S_n(f)}{a_{NG}^2}
\end{equation}
To compute $a_{NG}$, it is assumed that the main contributor is solar radiation pressure and that the spacecraft is at one astronomical unit (called $d_0$) from the Sun. The power carried by solar photons by surface unit at this distance is approximately equal to $P = 1.366 \times 10^3$~W~m$^{-2}$ \citep{willson2003secular}. Considering a ballistic coefficient equals to $C_B = 0.1$ m$^2$~kg$^{-1}$, which is the order of magnitude for Laplace mission \citep{biesbroek2008laplace}, the non-gravitational acceleration is equal to $a_{NG} = C_B P / c = 4.6 \times 10^{-7}$ m~s$^{-2}$ at one astronomical unit, where $c$ is the speed of light. Taking the minimum value of $S_n$, the requirement on $S_{\delta \theta}$ reads:
\begin{equation}
 \sqrt{S_{\delta \theta}(f)} \ll 1.3 \times 10^{-4} \ \mathrm{rad.Hz}^{-1/2}
\end{equation}
In the rest of the article, it will be assumed that this condition is verified and only the noise of MicroSTAR will be considered.

\subsection{Conditions for bias rejection}\label{subsection:condition}

The general approach presented above to solve the linear system does not give useful information on the non-gravitational acceleration or on the bias of the instrument. Since it is impossible to obtain the value of the unknown quantities at each sampling time, it is necessary to narrow the information retrieved from the data. A possibility is to look for the projection of the vectors $\mathbf{a_Y}$ and $\mathbf{a_Z}$ on a vector subspace (of dimension $p_a \leq N$) whose basis is made of the column of a matrix $V_a \in \mathcal{M}_{N,p_a}(\mathbb{R})$, which are supposed to be orthogonal for the usual scalar product on $\mathbb{R}^N$. As a result, the goal is to find the numerical values of $V_a' \mathbf{a_Y}$ and $V_a' \mathbf{a_Z}$ knowing $\mathbf{m_y}$ and $\mathbf{m_z}$ ($M'$ is the matrix transpose of $M$).
In this article, the choices of $V_a$ will allow retrieving the mean value of the acceleration without bias and  the slope of the acceleration over one modulation period. But other choices of $V_a$ can be made to retrieve for example sinusoidal variations of the signal.

Under the following four conditions on the bias, the angle~$\theta$ and the projection matrix $V_a$
\begin{equation}
  {V_a}' \Lambda_\nu \mathbf{b_\kappa} = 0, \ \mathrm{with} \ \nu\in\{c;s\} \ \mathrm{and} \ \kappa\in\{\mathbf{y};\mathbf{z}\},
  \label{eq:condition_demodulation}
\end{equation}
and assuming that $\delta k_{1y} = \delta k_{1z} = \delta k_{1}$, the unbiased values of the external signal can be recovered:
\begin{subequations}
 \begin{numcases}{}
  V_a'(1+\delta k_{1})  \mathbf{a_Y} =  V_a' \Lambda_c \mathbf{m_y} - V_a' \Lambda_s \mathbf{m_z} \\
  V_a'(1+\delta k_{1})  \mathbf{a_Z} =  V_a' \Lambda_s \mathbf{m_y} + V_a' \Lambda_c \mathbf{m_z}
 \end{numcases}
 \label{eq:a_simplified}
\end{subequations}
Calling $\mathbf{v}_k$ the $k$-th column of $V_a$, the conditions \eqref{eq:condition_demodulation} can be expressed in the frequency domain
\begin{equation}
 \left<\mathcal{F}_{\delta t}\{\mathbf{v}_k \cos(\mathbf{\theta})\},\mathcal{F}_{\delta t}\{\mathbf{b}_\kappa\}\right> = 0
\end{equation}
where $\mathcal{F}_{\delta t}$ is the discrete time Fourier transform and $\left<\cdot\right>$ is the usual scalar product. This equation means that the bias and the modulated signal must be orthogonal in the frequency domain.

It is a priori not possible to know whether conditions \eqref{eq:condition_demodulation} are fulfilled since the temporal evolution of the bias of the instrument is not controlled. However, as already mentioned, the bias corresponds deterministic low frequency variations. It is therefore possible to assume that $\mathbf{b_y}$ and $\mathbf{b_z}$ belongs to a vector subspace defined by the columns of $\hat{V}_b \in \mathcal{M}_{N,\hat{p}_b}(\mathbb{R})$ ($\hat{p}_b\leq N$). Given this hypothesis, conditions \eqref{eq:condition_demodulation} come down to
\begin{subequations}
  \begin{numcases}{}
    {V_a}' \Lambda_c \hat{V}_b =0 \\
    {V_a}' \Lambda_s \hat{V}_b =0
  \end{numcases}
  \label{eq:condition_demodulation_g}
\end{subequations}

The results presented in this section can be found by solving equation \eqref{eq:linear_system} with a modified least square method. The matrix $J$, which is unknown (because of the scale factors), is replaced by the matrix $\tilde{J} \in \mathcal{M}_{N,2(p_a+p_b)}$
\begin{equation}
  \tilde{J} =
  \begin{bmatrix}
      \Lambda_c V_a &  \Lambda_s V_a & V_b & 0 \\
    - \Lambda_s V_a &  \Lambda_c V_a & 0 & V_b
  \end{bmatrix}
\end{equation}
with $V_b \in \mathcal{M}_{N,p_b}(\mathbb{R})$. And it is assumed that $\mathbf{a_Y}$ and $\mathbf{a_Z}$ belong to the subspace generated by $\hat{V}_a \in \mathcal{M}_{N,\hat{p}_a}$ and that $\mathbf{b_y}$ and $\mathbf{b_z}$ belong to the subspace generated by $\hat{V}_b$. Section~\ref{subsection:GLS} will build on this approach.

\section{Unbiased measurements of non-gravitational acceleration}\label{section:measurement}

Based on the conditions \eqref{eq:condition_demodulation_g} required for a correct demodulation and given some assumptions on the matrices $V_a$ and $\hat{V}_b$, it is possible to design a calibration signal, i.e. a pattern for the angle $\theta$, which allows for completely removing the bias from the measurements.

\subsection{Choice of a calibration signal}\label{subsection:choice_signal}

The calibration signal looked for will be periodic, with a period called $\tau$. First, some practical concerns restrict the possible pattern. Because it can be assumed that rotating the accelerometer will induce vibrations and therefore spoil the measurements, the angle $\theta$ will have to be constant when the measurements are done. As result, calibration signals such that $\theta(t) = 2 \pi f t$, where $f$ is an angular frequency, are forbidden. Moreover, because the accelerometer may not be perfectly centered on the rotating plate, the rotation induces Coriolis and Centrifugal forces which spoil the signal. Finally, rotating constantly may lead to a quicker breakdown of the instrument.

Another practical concern, which appears if no slip ring is used, is about the wires between the accelerometer and the spacecraft. Because of them, it is not possible to rotate the accelerometer indefinitely. Therefore, the angle $\theta$ will have to stay in the interval $[0;2\pi]$.

To go further, it is necessary to be more specific on the matrices $V_a$ and $\hat{V}_b$. First, constant values of the non-gravitational acceleration during each modulation period will be looked for and the bias of the instrument will be supposed to be, for each period, an affine function of temperature. Therefore,
\begin{equation}
 V_a =
 \begin{bmatrix}
  \mathbf{1}_q &         & \mathbf{0}      \\
               &  \ddots &                 \\
  \mathbf{0}   &         & \mathbf{1}_q \\
 \end{bmatrix}
 \ \ \mathrm{and} \ \
 \hat{V}_b =
 \begin{bmatrix}
  \mathbf{1}_q    &         & \mathbf{0}   & \vdots \\
                  &  \ddots &              & \mathbf{T} \\
  \mathbf{0}      &         & \mathbf{1}_q & \vdots \\
 \end{bmatrix}
 \label{eq:def_V_1}
\end{equation}
where $\mathbf{1}_q$ is a matrix of $\mathcal{M}_{q,1}(\mathbb{R})$ whose coefficients are $1$, and $\mathbf{T}$ is a matrix of $\mathcal{M}_{N,1}(\mathbb{R})$ made of the values of the temperature at each sampling time. The integer $q$ is the number of sampling points in one period. It is assumed that $\tau$ and $f_s$ are such that $\tau f_s$ is an integer and $q = \tau f_s$. In this approach, the variation of temperature will be assumed to be driven by the heat generated by the rotating platform: at each rotation, heat is generated and induces a temporary increase of temperature.

\begin{figure}[ht]
 \begin{center}
  \subfigure[]{\includegraphics[width=0.35\linewidth]{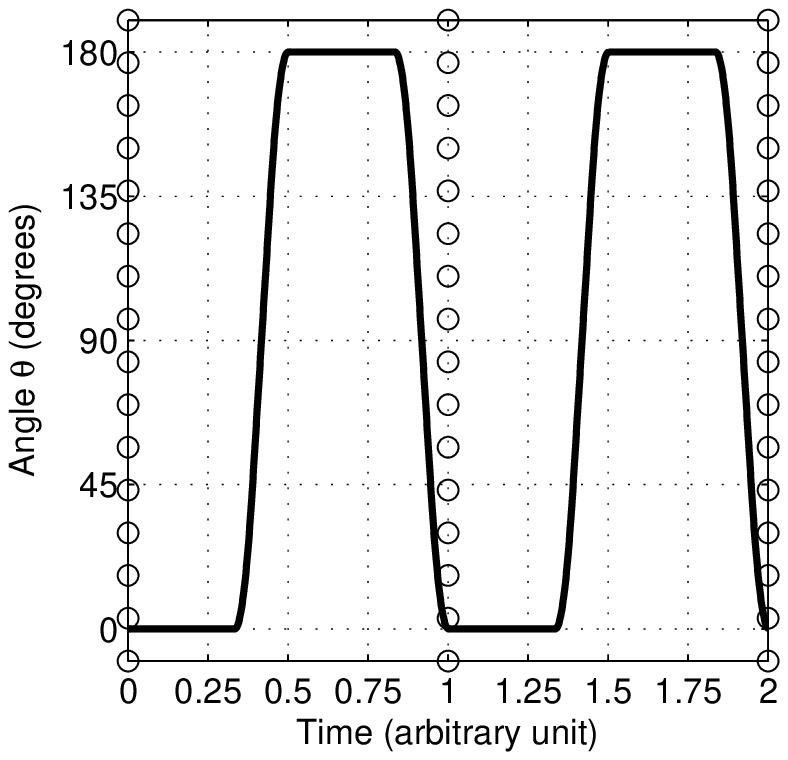}\label{fig:modulation_signal_a}}
  \hskip 30 pt
  \subfigure[]{\includegraphics[width=0.35\linewidth]{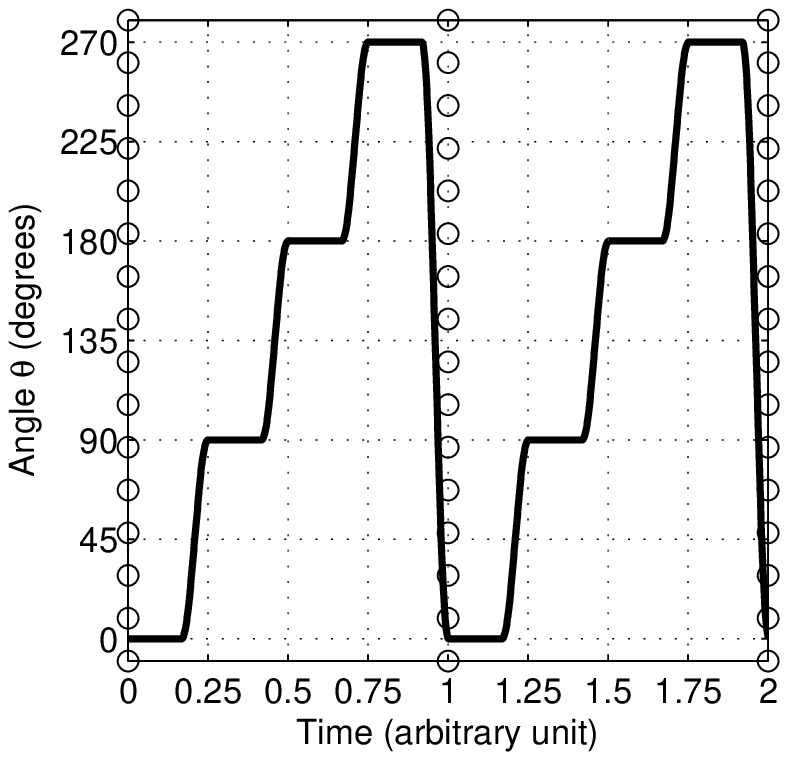}\label{fig:modulation_signal_b}}
  \caption{Example of two calibration signals $\theta(t)$ which fulfill the conditions given by equations \eqref{eq:condition_demodulation_g} for the matrices $V_a$ and $\hat{V}_b$ defined by equations \eqref{eq:def_V_1}. The rotating duration corresponds to 33.3 \% of the modulation period $\tau=1$~arbitrary unit. Two periods are represented, separated by circles ($\circ$).}
  \label{fig:modulation_signal}
 \end{center}
\end{figure}

Figure \ref{fig:modulation_signal} shows two examples of calibration signals which fulfill the conditions \eqref{eq:condition_demodulation_g} under the previous assumptions for the bias, the non-gravitational acceleration and the temperature. Let us consider the signal of Fig.~\ref{fig:modulation_signal_a} and go back to the assumptions made previously on the linear and quadratic factors\footnote{It was assumed that $k_{2y} = k_{2z} = 0$ and $\delta k_{1y} = \delta k_{1z}$.}. In Eq.~\eqref{eq:pos_pb}, the quadratic terms are constant because these two equalities are always true: $\sin(\theta)=0$ and $[\cos(\theta)]^2=1$. Therefore, the quadratic terms behave as a bias which will be separated from the non-gravitational acceleration. Concerning the assumption on the equality of the scale factors, it has to be noticed that ${\Lambda_c}^2 = \mathrm{Id}_N$ and $\Lambda_s = 0$.
Therefore, the derivation leading to Eq.~\eqref{eq:a_simplified} still hold without the assumption on the scale factors. On the contrary, for the signal of Fig.~\ref{fig:modulation_signal_b} as well as for any signal for which $\theta$ has values different from $0\char23$ and $180\char23$, these remarks on the scale and quadratic factors do not apply. As a conclusion, only signals for which measurements are made when $\theta = 0\char23$ and $\theta = 180\char23$ should be considered.

\begin{figure}[ht]
 \begin{center}
  \includegraphics[width=0.5\linewidth]{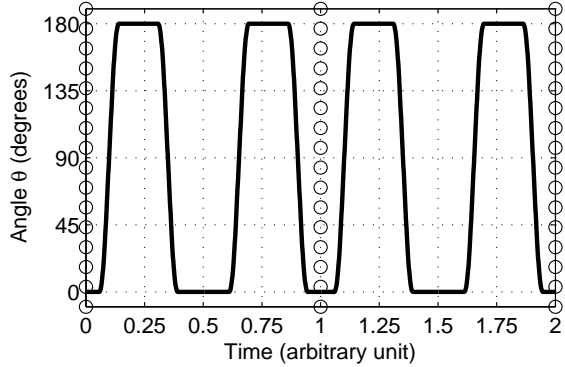}
  \caption{Example of a calibration signals $\theta(t)$ which fulfill the conditions given by equations \eqref{eq:condition_demodulation_g} for the matrices $V_a$ and $\hat{V}_b$ defined by equations \eqref{eq:def_V_2}. The rotating duration corresponds to 33.3 \% of the modulation period $\tau=1$~arbitrary unit. Two periods are represented, separated by circles ($\circ$). This signal is different from the one of Fig.~\ref{fig:modulation_signal_a} because it does not display a periodicity of 0.5~arbitrary unit.}
  \label{fig:modulation_signal_2}
 \end{center}
\end{figure}

The previous hypothesis made on $V_a$ and $\hat{V}_b$ allows to derive simple calibration signals. They are however restrictive because it is assumed that during a modulation period the signal and the bias are constant (with a temperature dependence for the bias). To go further, it is possible to design a calibration signal assuming that the bias is for each period an affine function of time but does not depend on temperature \citep{lenoir2011measuringSF2Aproceedings}. With this signal the mean and the slope of the non-gravitational acceleration on each modulation period will be recovered. In this case,
\begin{equation}
 V_a = \hat{V}_b =
 \begin{bmatrix}
  \mathbf{1}_q &         & \mathbf{0}   & \mathbf{t}_q &         & \mathbf{0}      \\
               &  \ddots &              &              &  \ddots &                 \\
  \mathbf{0}   &         & \mathbf{1}_q & \mathbf{0}   &         & \mathbf{t}_q \\
 \end{bmatrix}
 \label{eq:def_V_2}
\end{equation}
where $\mathbf{t}_q$ is a matrix of $\mathcal{M}_{q,1}(\mathbb{R})$ such that ${\mathbf{t}_q}_k = (k-q/2)\delta t$. Figure \ref{fig:modulation_signal_2} shows a calibration signal which fulfill conditions \eqref{eq:condition_demodulation_g}. Note that the remarks made on the scale and quadratics factors hold for this calibration signal. Contrary to the calibration signals of Fig.~\ref{fig:modulation_signal}, the pattern in this case depends on the masking time which is introduced in the following section. In the rest of this article this calibration signal will be used.

In case the bias of MicroSTAR does not belong to the subspace generated by $\hat{V}_b$, then the signal of interest is not perfectly recovered: it is spoiled by the quantities $V_a' \Lambda_\nu \mathbf{b_\kappa}$ ($\nu\in\{c;s\}$, $\kappa\in\{y;z\}$).

\subsection{Masking}\label{subsection:masking}

As mentioned in the previous paragraph, measurements made when the accelerometer is rotating are not considered for data reduction because they may be spoiled by unwanted signals. Therefore, during post-processing, the data acquired when the accelerometer is rotating must not be taken into account. This will be refered to as ``masking''. To introduce this masking feature in the signal processing, let consider the diagonal matrix $M \in \mathcal{M}_{N}(\mathbb{R})$ defined by: $M_{kk}=1$ if $\dot{\theta}_k=0$ and $\ddot{\theta}_k=0$, and $M_{kk}=0$ otherwise\footnote{$\dot{\theta}$ and $\ddot{\theta}$ are respectively the angular velocity and the angular acceleration of the rotating platform.}. Then in Eq.~\eqref{eq:a_simplified}, the matrix $V_a$ is replaced by $\tilde{V}_a = M V_a$.

The duration of masking is a key parameter in the precision of the post-processed quantities: the longer it is, the more data points are lost and the uncertainty increases (cf. section \ref{subsection:optimization}). The total duration of masking during one period is called $T_M$.

\section{Demodulated quantities}\label{section:demodulated-quantities}

The demodulation signals introduced in the previous section allows to retrieve unbiased measurements of the non-gravitational acceleration of the spacecraft. The focus will be now to characterize these post-processed quantities in term of uncertainty.

\subsection{Autocorrelation of the non-gravitational acceleration mean}\label{subsubsection:acceleration}

The calibration signal of Fig.~\ref{fig:modulation_signal_2} allows to recover affine variations of the external signal on each modulation period. In term of spacecraft navigation, the goal of the instrument is to measure the impact of non-gravitational forces on the dynamics of the spacecraft. And the variation of momentum during one modulation period $\overrightarrow{\Delta p_{NG}}$ of the spacecraft due to the non-gravitational forces $\overrightarrow{F_{NG}}$ is equal to the mean of the non gravitational forces times the modulation period:
\begin{equation}
 \overrightarrow{\Delta p_{NG}} = \int_{t_0}^{t_0+\tau} \overrightarrow{F_{NG}}(t) dt = \tau \left< \overrightarrow{F_{NG}} \right>_\tau
\end{equation}
where $t_0$ is an arbitrary time and $\left< \cdot \right>_\tau$ is the mean during a duration $\tau$.

As a result, only the mean values of the external signal are of interest, and the subsequent analysis will be restricted to the matrix $V_a$ defined by equation \eqref{eq:def_V_1}. Under the assumption introduced previously, the demodulated acceleration are defined by Eq.~\eqref{eq:a_simplified}. In order to have normalized quantities, it is necessary, as in the least square method, to multiply this equation on the left by $(\tilde{V}_a'\tilde{V}_a)^{-1}$. Under the assumptions considered here, this matrix is diagonal with all the coefficients equal to $q = |\mathbf{\tilde{v}}_i|^2$, where $\mathbf{\tilde{v}}_i \in \mathcal{M}_{N,1}(\mathbb{R})$ is $i$th column of the matrix $\tilde{V}_a$. $q$ is independent of the index $i$.

Let us call $\mathbf{c}_i = \Lambda_c \mathbf{\tilde{v}}_i \in \mathcal{M}_{N,1}(\mathbb{R})$ the $i$th column of the matrix $\Lambda_c \tilde{V}_a$, $\widehat{a_Y}_i = (1+\delta k_{1y}) \mathbf{\tilde{v}}_i'\mathbf{a_Y}/q$ the $i$th component of the column vector $(1+\delta k_{1y}) (\tilde{V}_a'\tilde{V}_a)^{-1} \tilde{V}_a'\mathbf{a_Y}$, and $\widehat{a_Z}_i = (1+\delta k_{1z}) \mathbf{\tilde{v}}_i'\mathbf{a_Z}/q$ the $i$th component of the column vector $(1+\delta k_{1z}) (\tilde{V}_a'\tilde{V}_a)^{-1} \tilde{V}_a'\mathbf{a_Z}$.

The quantities $\widehat{a_Y}_i$ and $\widehat{a_Z}_i$ are the means of the non-gravitational acceleration of the spacecraft for the modulation period $i$ along the axes $Y$ and $Z$ respectively. Under the assumption made earlier, the accuracy of the measurements is perfect, i.e. their expected values is equal to the true values. Concerning the precision, assuming that $n_y$ and $n_z$ are independent and have the same power spectrum density, $S_n$ defined by equation \eqref{eq:acc_noise}, the covariances between the post-processed quantities are
\begin{equation}
 \mathrm{Cov}(\widehat{a_Y}_i,\widehat{a_Y}_j) = \mathrm{Cov}(\widehat{a_Z}_i,\widehat{a_Z}_j) = \int_{-\frac{1}{2\delta t}}^{\frac{1}{2\delta t}} S_n(f) \left( \frac{\mathcal{F}_{\delta t}\{\mathbf{c}_i\}(f)\overline{\mathcal{F}_{\delta t}\{\mathbf{c}_j\}(f)}}{q . \delta t^2} \right) df
  \label{eq:covariance}
\end{equation}
and
\begin{equation}
 \mathrm{Cov}(\widehat{a_Y}_i,\widehat{a_Z}_j) = 0
\end{equation}
\begin{figure}[ht]
 \begin{center}
  \subfigure[]{\includegraphics[width=0.35\linewidth]{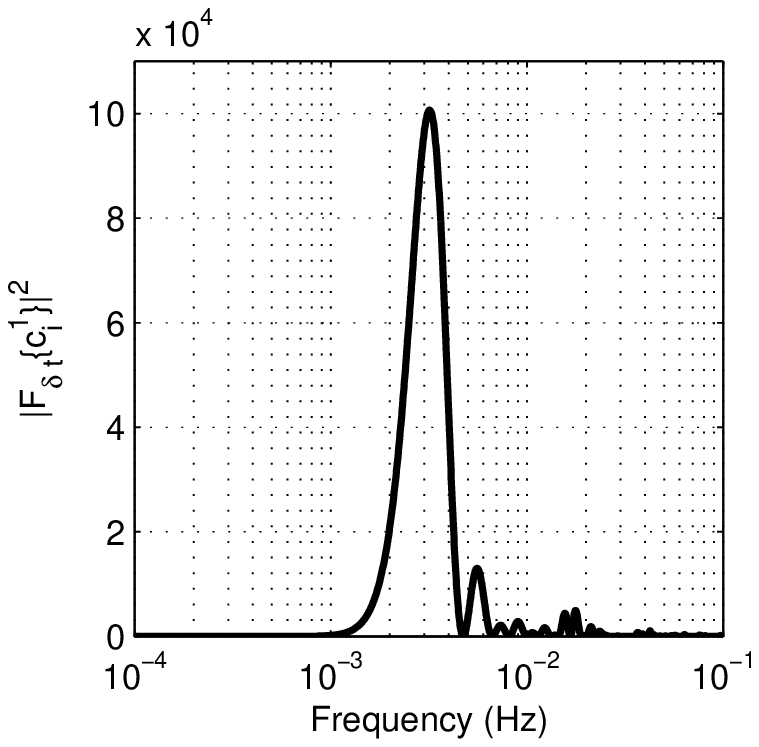}\label{fig:DTFT_a}}
  \hskip 30 pt
  \subfigure[]{\includegraphics[width=0.35\linewidth]{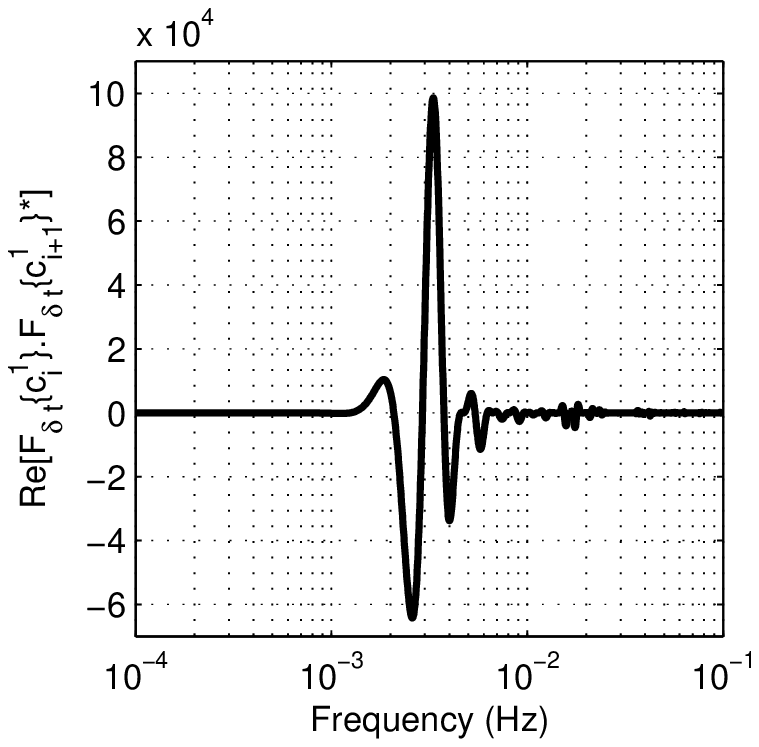}\label{fig:DTFT_b}}
  \caption{$\mathrm{Re}\left[\mathcal{F}_{\delta t}\{\mathbf{c}_i\}\overline{\mathcal{F}_{\delta t}\{\mathbf{c}_j\}}\right]$ for the calibration signal of Fig.~\ref{fig:modulation_signal_2} with a modulation period $\tau$ of 600 s, a masking time $T_M$ of 200 s and a sampling frequency $f_s$ of 1 Hz: \subref{fig:DTFT_a} $i=j$, \subref{fig:DTFT_b} $j = i+1$. For figure~\subref{fig:DTFT_a}, the peak is approximately at the frequency $2/\tau$ and its frequency width is approximately $1/\tau$.}
  \label{fig:DTFT}
 \end{center}
\end{figure}
The result given by equation \eqref{eq:covariance} is true only if the signal has been filtered before digitization by a perfect low-pass filter with a cut-off frequency of $f_s/2$ so as to avoid aliasing.

According to Fig.~\ref{fig:DTFT}, the integral of equation \eqref{eq:covariance} select the noise power spectrum density at the frequency $1/\tau$ and approximately integrate it on an bandwidth $1/\tau$ for $i=j$. In order to minimize the absolute value of the covariance, it is therefore necessary to select the noise at the frequencies where it is minimum, i.e. for $f\in [10^{-2};2\times10^{-1}]$ Hz, which correspond approximately to modulation period between 5 s and 100 s.
Too short modulation periods are impossible to implement in practice. Therefore, in the following, a modulation period equal to $10$ min will be considered.

\begin{figure}[ht]
 \begin{center}
  \includegraphics[width=0.5\linewidth]{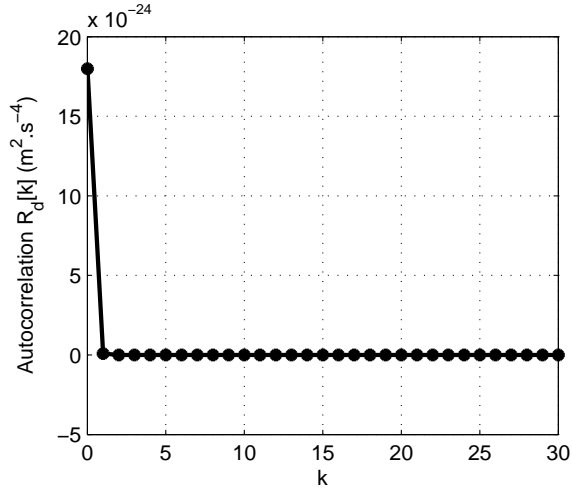}
  \caption{Autocorrelation function $R_d[k]$ defined by equation \eqref{eq:auto_dem_a} for the calibration signal of Fig.~\ref{fig:modulation_signal_2} with a modulation period $\tau$ of 600 s, a masking time $T_M$ of 200 s and a sampling frequency $f_s$ of 1 Hz.}
  \label{fig:auto_dem_a}
 \end{center}
\end{figure}

The main interest of the demodulation process is to know the mean acceleration over a modulation period $\tau$. This process gives birth to two new discrete-time quantities $\widehat{a_Y}_i$ and $\widehat{a_Z}_i$ indexed formally by $i\in\mathbb{Z}$. It is possible to introduce the autocorrelation function $R_d[k]$ which is the same for both quantities and which is defined by
\begin{equation}
 R_d[k] = \mathrm{Cov}(\widehat{a_Y}_{i+k},\widehat{a_Y}_i) = \mathrm{Cov}(\widehat{a_Z}_{i+k},\widehat{a_Z}_i).
 \label{eq:auto_dem_a}
\end{equation}
Fig.~\ref{fig:auto_dem_a} shows that the autocorrelation function is close to the one of a white noise. This means that the post-processed quantities are approximately independent. In term of power spectrum density, this corresponds to a level of $10^{-10}$~m~s$^{-2}$~Hz$^{-1/2}$ with a cut-off frequency equal to $8.3\times10^{-4}$~Hz. Since the uncertainty on the demodulated accelerations is known and characterized, it is now possible to use them to gain more information on the non-gravitational accelerations.

\begin{figure}[ht]
 \begin{center}
  \includegraphics[width=0.5\linewidth]{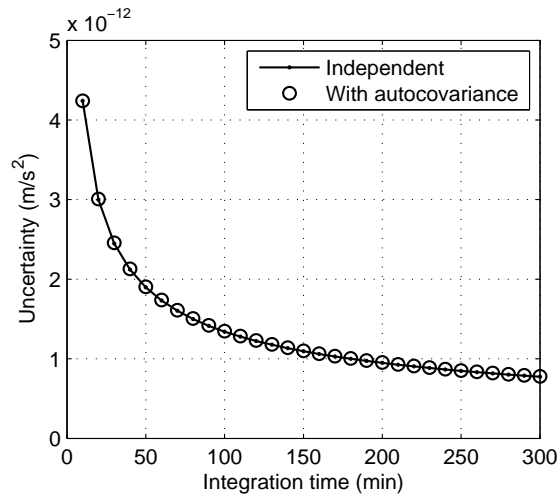}
  \caption{Uncertainty on the mean acceleration as a function of the integration time assuming that the post-processed quantities are independent or have the autocorrelation function plotted in Fig.~\ref{fig:auto_dem_a}. The plot is for the calibration signal of Fig.~\ref{fig:modulation_signal_2} with a modulation period $\tau$ of 600~s, a masking time $T_M$ of 200~s and a sampling frequency $f_s$ of 1~Hz. Its shows that the independence approximation is correct. For an integration time of 3 hours, the uncertainty is equal to 1~pm.s$^{-2}$ on the mean acceleration.}
  \label{fig:uncertainty_mean}
 \end{center}
\end{figure}

\subsection{Further characterization of the non-gravitational acceleration}

In order to increase the precision, it may be interesting to know the mean acceleration over periods of time longer than the modulation period. To do so, one needs to average the demodulated accelerations over the period of time of interest. Figure \ref{fig:uncertainty_mean} shows the uncertainty on the mean acceleration for different integration time. As the noise is nearly white, the uncertainty on the mean decreases as $1/\sqrt{T}$ where $T$ is the integration time.

It is also possible to look for sinusoidal variations of the non gravitational acceleration with a known frequency $f^*$. The goal is to find the coefficients $\alpha$ and $\beta$ of the time varying signal $\alpha \cos(2\pi f^* t) + \beta \sin(2\pi f^* t)$ using the values of the non-gravitational acceleration for each modulation period. According to the Nyquist--Shannon theorem, it is not possible to recover sinusoidal variations at frequencies higher than half the sampling frequency, i.e. $f^* \geq 1/(2 \tau)$. Conversely, when $1/(\tau f^* )$ becomes too large, the uncertainty diverges. The value for which this happens depends on the number of post-processed points used to fit the sinusoidal variation: the more points are used, the easier it is to fit low frequency signals.
For $\tau=10$~min and $f^*$ the frequency related to the revolution period of the Earth\footnote{Because the instrument will be used for spacecraft tracking, one may be interesting in sinusoidal variations at the revolution period of the Earth. These variations are also of interest for the fundamental physics objectives discussed in the introduction.}, the frequency ratio is $1/(\tau f^* )= 144$. In this particular configuration, one obtains with 60 points and for a modulation period of 10~minutes (which corresponds to 10~hours of measurement):
\begin{subequations}
 \begin{numcases}{}
  \sqrt{\mathrm{Cov}(\alpha,\alpha)} = 8.7 \times 10^{-13} \ \mathrm{m.s}^{-2} \\
  \sqrt{\mathrm{Cov}(\beta,\beta)} = 7.3 \times 10^{-13} \ \mathrm{m.s}^{-2} \\
  \sqrt{-\mathrm{Cov}(\alpha,\beta)} = 2.4 \times 10^{-13} \ \mathrm{m.s}^{-2}
 \end{numcases}
\end{subequations}
These values show that it is possible to obtain, in this configuration, the amplitude of the sinusoid with a precision better than 1~pm~s$^{-2}$.

\subsection{Generalized least square/optimal filtering}\label{subsection:GLS}

In section \ref{subsection:condition}, it was mentioned that the process described until now corresponds to a least square (LS) method. This method provides estimates with a minimum variance only when the noise is white. However, the noise of MicroSTAR does not fall in this category. The generalized least square (GLS) method \citep{cornillon2007regression} provides an estimate with minimum variance whatever the measurement noise is.

This method is similar to the optimal filtering technique \cite[p. 325]{papoulis1977signal}: the first one is expressed in the time domain whereas the second one is express in the frequency domain. It is possible to express the components of the inverse of the covariance matrix $V_{GLS} = (\tilde{J}' \Omega^{-1} \tilde{J})^{-1}$ using the power spectrum density of the noise instead of its covariance matrix: if $\mathbf{v}$ and $\mathbf{w}$ are two column vectors, then
\begin{equation}
 \mathbf{v}' \Omega^{-1} \mathbf{w} = \int_{-\frac{1}{2\delta t}}^{\frac{1}{2\delta t}} \frac{1}{S_n(f)} \mathcal{F}_{\delta t}\{\mathbf{v}\}(f)\overline{\mathcal{F}_{\delta t}\{\mathbf{w}\}(f)} df
\end{equation}

One may process the data form the accelerometer using the generalized least square method. However, in the specific case of the problem studied here, the gain is rather small. Indeed, Fig.~\ref{fig:DTFT} shows that, for the calibration signal considered, the Discrete Time Fourier Transform is peaked around the frequency $2/\tau$ and the noise PSD does not vary much on the interval $[10^{-3};10^{-1}]$~Hz. As a result, using Parseval theorem,
\begin{equation}
 \left( \int_{-\frac{1}{2\delta t}}^{\frac{1}{2\delta t}} \frac{1}{S_n(f)} \left|\mathcal{F}_{\delta t}\{\mathbf{c_i}\}(f)\right|^2 df \right)^{-1} \approx \frac{1}{\tau} S_n\left(\frac{2}{\tau}\right) \approx \int_{-\frac{1}{2\delta t}}^{\frac{1}{2\delta t}} S_n(f) \left|\frac{\mathcal{F}_{\delta t}\{\mathbf{c_i}\}(f)}{q . \delta t} \right|^2 df
\end{equation}
Therefore, the autocorrelation function plotted in Fig.~\ref{fig:auto_dem_a} is nearly the same as the one obtained with the GLS approach: the autocorrelation function obtained with the GLS approach is similar to the one of a Gaussian noise and its value for $k=0$ is $1.71\times 10^{-23}$~m$^2$~s$^{-4}$ instead of $1.84\times 10^{-23}$~m$^2$~s$^{-4}$ for Fig.~\ref{fig:auto_dem_a}. The difference in the level of precision on the post-processed quantities is also visible on Fig.~\ref{fig:uncertainty_optimization} in section~\ref{subsection:optimization}.

\subsection{Optimization of the masking time and calibration period}\label{subsection:optimization}

\begin{figure}[ht]
 \begin{center}
  \subfigure[]{\includegraphics[width=0.48\linewidth]{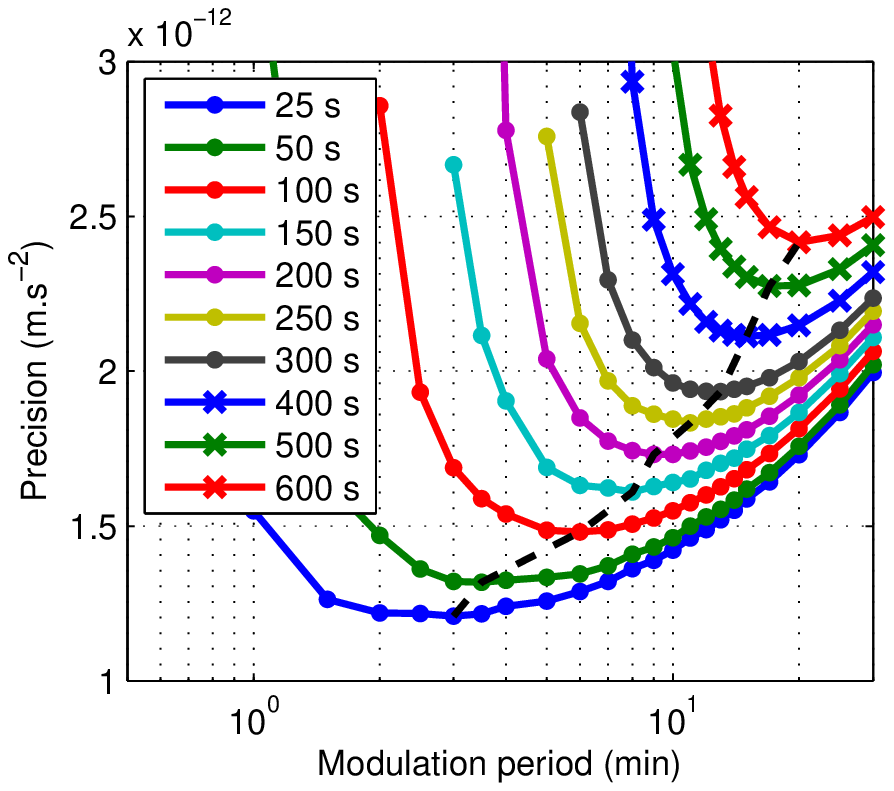}\label{fig:uncertainty_optimization_a}}
  \hskip 10 pt
  \subfigure[]{\includegraphics[width=0.48\linewidth]{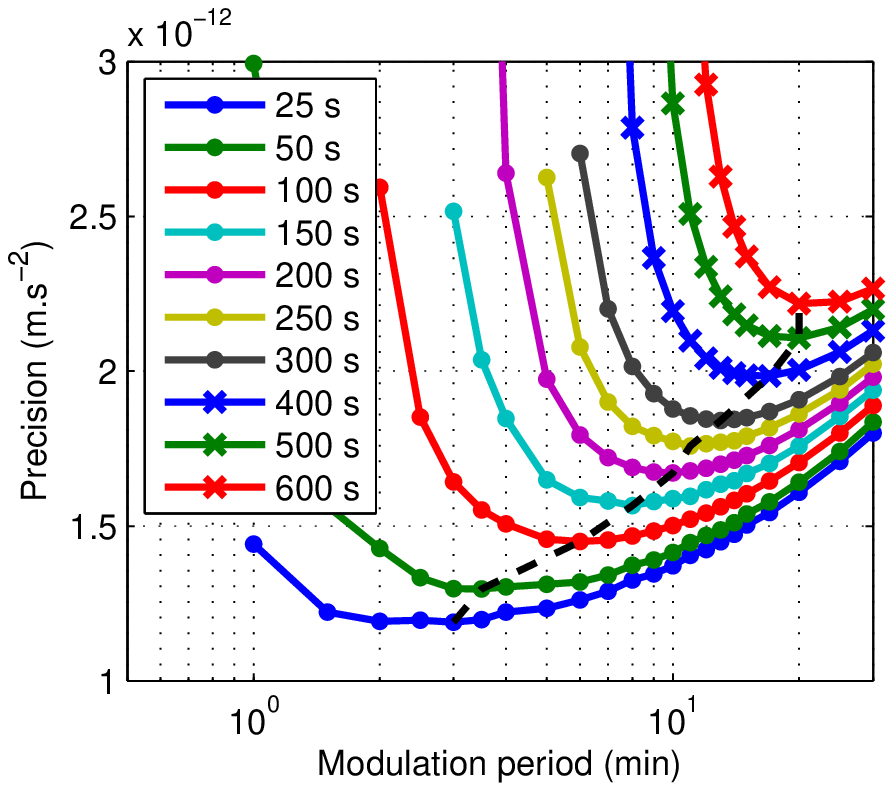}\label{fig:uncertainty_optimization_b}}
  \caption{Uncertainty on the demodulated acceleration for an integration time of one hour and for the calibration signal of Fig.~\ref{fig:modulation_signal_2} as a function of the modulation period. The plots are parametrized by the masking time. The dash line shows the modulation period which gives the minimum uncertainty for each masking time. Figure~\subref{fig:uncertainty_optimization_a} is obtained using the least squares method to process the data whereas figure~\subref{fig:uncertainty_optimization_b} is obtained with the generalized least squares method.}
  \label{fig:uncertainty_optimization}
 \end{center}
\end{figure}

Until now, only one modulation period ($\tau=10$ min) and one masking time ($T_M=200$ s) have been considered. But since their value impact the uncertainty on the demodulated accelerations, it is legitimate to choose these values such that the uncertainty is minimized.

Figure \ref{fig:uncertainty_optimization} shows the uncertainty on the demodulated acceleration for an integration time of one hour and for the LS (a) and GLS (b) methods. This value is computed by taking the numerical value of $R_d[0]$ and by multiplying it by $\sqrt{\tau/T}$, where $\tau$ is the modulation period and $T = 1$~hour. This assumes that the demodulated accelerations are independent, which has been shown to be true. As what was already said, the smaller the masking time is, the smaller the uncertainty is. However, instrumental constraints do not allow to rotate MicroSTAR too fast. For a given masking time, Fig.~\ref{fig:uncertainty_optimization} gives the optimal modulation period. For example, it shows that the set of parameters used until now ($\tau=10$~min and $T_M=200$~s) is ``optimal'' for the GLS approach, i.e. $\tau=10$ min gives the minimum uncertainty for a masking time of 200~s.

\section{Conclusion}\label{section:conclusion}

The Gravity Advanced Package, developed to improve orbit reconstruction of interplanetary probes in order to test General Relativity, relies on a technological progress with allows using an electrostatic accelerometer to make measurements with no bias. Indeed, the addition of a rotating platform allows modulating the non-gravitational acceleration while keeping the bias at low frequencies. The data acquired need to be processed in order to obtain the measurement with no bias. This data processing was the topic of this article.

The first result obtained was conditions under which the bias is completely removed from the signal of interest. These conditions allowed designing calibration signals, i.e. a time-pattern for the rejection angle. Then the uncertainties on the unbiased non-gravitational acceleration were computed. It was shown that it is possible to recover the mean acceleration over each period of modulation and to have access to sinusoidal variations of this accelerations with some restriction on the pulsation of the signal. Finally, a method was presented to optimize the modulation period and the masking time so as to reach the minimum uncertainty.

It has been shown that several parameters influence the precision on the post-processed quantities: the modulation time, the masking time and the integration time. It is possible to choose a set of parameters, which are technologically speaking reasonable, leading to precision below 1~pm~s$^{-2}$ on mean quantities as well as on the amplitude of sinusoidal variations. This precision is expected to improve orbit reconstruction significantly.

\section*{Acknowledgements}\label{section:acknowledgements}

The authors are grateful to CNES (Centre National d'\'Etudes Spatiales) for its financial support.




\end{document}